# A Usefulness-based Approach for Measuring the Local and Global Effect of IIR Services

Daniel Hienert and Peter Mutschke
GESIS – Leibniz Institute for the Social Sciences
Cologne, Germany
{firstname.lastname}@gesis.org

## ABSTRACT
In Interactive Information Retrieval (IIR) different services such as search term suggestion can support users in their search process. The applicability and performance of such services is either measured with different user-centered studies (like usability tests or laboratory experiments) or, in the context of IR, with their contribution to measures like precision and recall. However, each evaluation methodology has its certain disadvantages. For example, user-centered experiments are often costly and small-scaled; IR experiments rely on relevance assessments and measure only relevance of documents. In this work we operationalize the usefulness model of Cole et al. (2009) on the level of system support to measure not only the local effect of an IR service, but the impact it has on the whole search process. We therefore use a log-based evaluation approach which models user interactions within sessions with positive signals and apply it for the case of a search term suggestion service. We found that the usage of the service significantly often implicates the occurrence of positive signals during the following session steps.

## 1. INTRODUCTION
There are numerous services which support users in their information seeking process such as term or query suggestions, personalization, faceted navigation, relevance feedback, visual representations, re-ranking mechanisms, browsing facilities or related items. Different methodologies are available to evaluate the applicability, usability, effectiveness or performance of such services. User-centered studies aim to identify usability or interaction problems which are based on the user's interaction with the system. This reveals the human's view on an IR service [cp. 20]. In classical IR the focus is on document relevancy which is measured by variables such as precision and recall on the basis of available relevance assessments. Accordingly, the influence of supporting services is measured by their positive impact to these traditional measures.

As a novel approach Cole et al. [4] introduced the notion of *usefulness* as a general criterion of "how well the user is able to achieve his goal" in the system under study. The authors present an IIR evaluation model which asks for usefulness on three different levels: a) the entire information seeking episode and the leading task, b) each interaction and its contribution to the leading task and c) the system support toward the goal of each interaction and of each information seeking strategy (ISS). This perspective takes "both task success and the value of support given over the entire information seeking episode" [4] into account. This represents a novel paradigm in IR evaluation insofar as it expands the perspective to the entire search process instead of just evaluating single search results with respect to relevancy [cp. 5]. However, it remains difficult in a complex environment to answer these questions, especially if the user's task and subtasks remain unclear. Moreover, there is still a lack of computational methods that can be applied to evaluate interactive IR systems.

In this work we try to operationalize the usefulness model on a local and global level in the form of a computational model that can be applied in large-scale evaluation studies. Following the different evaluation levels described by [4] we focus on the usefulness of a single service of an IR system for supporting a single ISS (local usefulness of the service) as well as its contribution to the usefulness of the entire system in accomplishing the user's information seeking goal (global usefulness of the service). Following [4] which introduces usefulness as a concept "suited to interaction measurement" we use a log-based approach focusing on interaction events that indicate a positive impact of the service. We then apply this evaluation methodology to the case of a search term suggestion service and discuss the results.

## 2. BACKGROUND AND RELATED WORK
### 2.1 Evaluation Models
The measurement of document relevancy with evaluation initiatives such as TREC [31] has been the predominant evaluation methodology over the last decades. However, the field of IR opens to a more holistic view of the search and information process and puts the user in context. Ingwersen & Järvelin [13] present an evaluation model that asks beside the (a) IR context, for the (b) seeking-, (c) work task and (d) socio-organizational/cultural context. That automatically leads to other evaluation criteria such as usability, quality of information process, quality of information and work process/results, socio-cognitive relevance and quality of work task result.

The usefulness model as a holistic evaluation model [cp. 4] assumes a *problematic situation* a user has by lacking knowledge about a topic. The general information seeking *goal/task* is then to achieve this knowledge. This overall goal can be subdivided into several *sub-goals*, each described by an *information interaction* to achieve the subgoal and, to this end, the general goal, e.g. collecting information, learning about the material or comparing results. Therefore, each information seeking episode can be seen as a sequence of *information seeking strategies* (IIS, [3]), e.g. querying, receiving results or evaluating documents which the IR system can support. Accordingly, usefulness can be measured on three levels as mentioned above.

This makes the model a good starting point for IIR evaluation because it describes the information seeking process from the user's point of view and how the IR system can support it. For the broad application of the model there are some challenges: (1) the users' overall goal and sub goals are often unknown in real world applications, (2) until now it has not been shown how the theoretical model of usefulness can be transformed into practice and how it can be operationalized.

As regards (1), it must be pointed out that the usefulness model is very much designed around precise knowledge about the leading goal and the following tasks of a particular user [34]. However, the overall task and sub tasks are often only available in a laboratory setting where evaluation studies are conducted in direct contact with users. In real-world systems, in contrast, knowledge about tasks is hard to collect. One possible solution is to explicitly ask the user for the task by some system dialogues, another is the extraction of tasks from log files by clustering search queries from web search engines [23,32]. The task-based session then contains all search queries for a particular search intention. However, the overall goal and task is still missing, especially with more complex and longitudinal information problems as in IIR.

As regards (2), Cole et al. [4] provide a non-exclusive list of questions at the different levels, such as "How *useful* were suggested queries/terms for formulating queries?" or "How *well* does the system support evaluation of retrieved documents?". The intention of Cole et al. [4] is to let the user give the answers within user studies. However, user studies are often small-scaled and very much specific to a particular system. In a large-scale evaluation setting, however, which needs a computational model, most of the evaluation questions proposed by [4] are hard to answer since adjectives such as "useful" or "well" are hard to capture by computational measures.

The central research question of our paper therefore is: "How can usefulness of a particular IR service under study be approximated in the form of log data based measures?" The availability of a reliable approach for this would allow large-scale experiments and the application in very different contexts and IR systems.

### 2.2 Evaluation Methodologies

For the field of Interactive Information Retrieval (IIR) Kelly [20] gives a good overview of existing and established evaluation methodologies and measures. She proposes a research continuum which has on the one side TREC-style studies which build the system focus on IIR evaluation and on the other side the observation of information-seeking in context which build the most human focus on IIR evaluation. The archetypical IIR evaluation study is represented by the TREC Interactive Track. Seen from there log analysis studies are situated one step towards the system focus. According to Kelly "search engine logs look at queries, search results and click-through behavior" [20] and log analysis is more descriptive than explanatory, also "it is possible to model user behavior and interactions for certain situations" [20].

For the basic possibilities and limitations of search log analysis Jansen [14] gives a good overview. Log analysis can identify trends and typical interactions, but cannot record the user's perception of the task, the underlying information need or the underlying situation and context of the search. A review of log analysis literature is given in [1]. The authors distinguish explicitly between Web search engine log analysis (WSE) and digital library log analysis (DLS) as in WSE the retrieved documents are web pages and in DLS documents with a quality maintained by professionals. Additionally, in DLS document collections are mostly organized and structured by a knowledge organization system and users in DL search are much more specific around a community of a domain or a certain topic.

### 2.3 Interaction Measurement

Interaction measurement as a methodic approach to solve IR problems is in line with current works addressing whole user sessions and multiple sessions. For example, Wildemuth [33] examined search tactics behavior of medical students searching a database in microbiology. She found patterns of search tactics where users added and deleted concepts to their search queries and shows that domain knowledge influences search tactics behavior. Jansen et al. [15] found analog to prior results that in web search main transition patterns are generalization and specialization. Additionally, different measures have been found as signals for session behavior. For example, Fox et al. [7] found in a user study that a combination of click-through, time spending on the search result page and how a user exited a result of a search session correlates best with user satisfaction. Liu et al. [22] identified three main behavioral measures as important for document usefulness in a laboratory experiment: dwell time on documents, the number of times a page has been visited during a session and the timespan before the first click after a query is issued. Predictive models have then been applied to the TREC 2011 Session Track and showed improvement over the baseline by using pseudo relevance feedback on the last queries in each session. Azzopardi [2] suggested different effectiveness measures for IR systems based on a stream-based view of documents in the IR process including a window-based approach. Thomas [29] uses positive and negative signals of web sessions such as session duration or scrolling events to determine if users are struggling. Navigation patterns that correlate with these signals can then help website authors to reveal navigational problems. Kelly [21] gives an overview of related work which utilizes implicit feedback from users, mainly applied for query expansion or user profiling. Implicit feedback is given by user behavior such as viewing, printing or quoting a document. Zhang & Kamps [36] for example used email correspondence between archivists and users which reference documents for specific topics as ground-truth. For the approach of click-through data, it is assumed that a document has certain relevance if the user clicks on it. Joachims et al. [16,17] analyzed click-through data as implicit feedback in web search and found that on average click data is reasonable accurate but biased by the trust in the retrieval function and the quality of the result set. Kamps et al. [18] compared click-through and user judgements on the base of different test collections and manually created and assessed topics. They found that in their comparison the agreement is only small and have some biases. For example the number of relevant documents for a topic depending on the test collection can differ.

### 2.4 Search Term Suggestion

Typically there is a gap between the user's natural language and the vocabulary an information system uses to index its documents which is described as the "vocabulary problem" [8]. Knowledge organization systems (KOS) such as thesauri, classifications and ontologies contain knowledge structures which can help improve the search process, for instance by expanding the search query (e.g. [35]) with near-by concepts for better retrieval results. On the user interface, users can be supported with a list of query or term suggestions. Search term recommenders today are widely implemented, from web search engines to e-commerce platforms. Terms proposed to the user can derive from a number of different sources. Efthimiadis [6] distinguishes between (1) collection dependent and independent knowledge structures such as thesauri and (2) knowledge from search results. Vechtomova [30], for example compared two approaches based on a co-occurrence analysis on the entire document collection and on a subset from the retrieved results and found that the local approach performed clearly better for query expansion. But also other sources such as query logs [12] have been used as vocabulary for term recommendation. Schatz et al. [28] compared term recommenders based on a subject thesauri and from a co-occurrence list. They conclude that a combination of both in one interface with multiple views can be advantageous for users to choose recommendations from multiple sources. Nowadays, term suggestion services are used in a lot of information systems especially on commercial platforms, but larger digital libraries such as the ACM Digital Library or Google Scholar are still struggling offering them (compare [25] for a short list). One of the reasons can be the broad content or a missing knowledge organization system to structure it.

The use of search term recommenders can improve the retrieval performance in the sense of document relevancy. So, it has been shown that query expansion based on a local subset of documents from the result list [26,35] or discipline-specific query expansion [24,27] can result in significantly better results. Thus, term suggestion services in digital libraries (especially in domain-specific ones with organized content) seem to be useful insofar as they can help to suggest the user query terms, titles, authors, journals and so on. However, beyond document relevancy, it is difficult to measure which effect an IR system or a particular service of the system has on the entire search or interaction process.

## 3. APPROACH

In contrast to many other measures, which evaluate single elements of a retrieval system (such as the quality of its ranking), usefulness aims at evaluating the degree to which the system under study helps the user in solving his/her information seeking problem. This includes the quality of the entire search process consisting of functionalities beyond pure searching offered by many digital libraries, such as navigating through link structures, structuring, sharing, storing and exporting information, which broaden the amount of possible user-system interactions. Given this, user-system-interactions lead to valuable data for a better understanding of user needs and information behavior. Following [4] usefulness "is suited to interaction measurements". Thus, we suppose that by a particular analysis of interaction measurement usefulness can be approximated beside from relevancy.

Cole et al. [4] consider usefulness on the level of the entire "information seeking episode" as well as on the level of each single ISS (given by an interaction) and its contribution to achieve the leading information seeking goal. Accordingly, we differentiate between local and global usefulness of the service that implements the ISS in question. Following the overview of Kelly [21] and also Thomas [29] which describe implicit feedback as indicators of user preference, user satisfaction and interests, an intuitive notion of local usefulness is the amount of positive choices of a particular ISS, given by invoking a single interaction such as the selection of a term provided by a term suggestion service. Therefore, we utilize the frequency of interactions that stand for the particular service under study to define *local usefulness* as the percentage of the services usage in all search processes. This is basically a usage-based notion of usefulness providing a clue of how useful the service is considered by users to achieve a certain sub goal (such as selecting a proper search term by the help of a search term recommender), and it is a local measure since it refers to the *current* phase of the search process: The more often the service has been used, the greater its (expected) usefulness for supporting the user on the local level of a single ISS.

On the level of the entire search session we can then ask for the degree to which the use of the service contributes to successfully accomplish the leading information seeking goal. An approach for estimating global usefulness is to count the amount of positive signals emanating from the (local) use of the service in question. Thus, we define *global usefulness* as the degree to which the use of a certain service on the local level leads to positive signals of search success in the succeeding phase of the search session. In contrast to local usefulness this is a success-based notion of usefulness, and it is a global measure since it refers to the entire search session: The more often the service in question leads to positive signals in a later phase of the session, the greater its usefulness in supporting the user on the level of the entire information seeking episode.

Formally, we define the *retrieval system* $R$ to consist of a set $D$ of *documents*, a set $E$ of possible *interaction events* and a set $U$ of different *users*:

$R = (D, E, U)$, where $D = \{d_1, \ldots, d_n\}$, $E = \{e_1, \ldots, e_n\}$ and $U = \{u_1, \ldots, u_n\}$

A *search process* $p$ is a sequence of search events $e \in E$ invoked by a user $u \in U$, starting with a start search event, e.g. $enter\_search\_term$ and ending with either a terminal event of the session, such as $logout$, or the last event preceding a new start search event which indicates the start of a new search process:

$p_u = (enter\_search\_term, e_2, \ldots, e_n)$

The explicit usage of a particular IR service by the user (e.g. choosing a term from a recommender) is indicated by the dedicated event $e_{IRService} \in E$.

Success events $SE \subset E$ are a subset of events indicating positive signals of success in a search process, e.g.

$SE_i = \{print\_record, export\_record, bookmark\_record\}$

A window of events $w(n)$ is a sequence of interaction events, starting with a particular initial event $e_{Initial} \in E$, followed by $n$ succeeding events:

$w(n) := (e_{Initial}, \ldots, e_n)$

$e_{IRService}^+(w(n))$ indicates search success in terms of incidence of positive signals following the use of the IR service in question. The function returns 1 if the usage of the IR service is followed by at least one success event within a window of $n$ succeeding events (for $e_{Initial} = e_{IRService}$), otherwise 0. However, a value of 1 does not mean that the use of the IR service causes the positive signal. But it points to the co-incidence of the two events in question during the search process. Our intention is to enable comparisons between different services as well as different searches with/without usage of a particular service as regards their effect on search success.

Similarly, $e_{Search}^+(w(n))$ is 1 if a search is followed by at least one success event in a window of $n$ succeeding events, otherwise 0.

The *local usefulness* of an IR service is then defined as the ratio of the count of all IR service usages to the number of all search processes:

$$LocalUsefulness\ (IRService) = \frac{\sum (e_{IRService})}{|p|}$$

The *global usefulness* of an IR service is defined as the ratio of the count of all IR service usages followed by a positive signal within a window of $n$ succeeding events to the number of all usages of the IR service.

$$GlobalUsefulness\ (IRService) = \frac{\sum (e_{IRService}^+(w(n)))}{|e_{IRService}|}$$

Both metrics provide numbers in the range [0:1].

This can be compared to the global usefulness of a search without the usage of the IR service which is defined as the ratio of the count of search events followed by a positive signal in a window of $n$ succeeding events to all search events:

$$GlobalUsefulness\,(Search) = \frac{\sum\,(e^+_{Search}(w(n)))}{|e_{Search}|}$$

The values of global usefulness for *IRService* and *Search* can then finally be compared in order to find the smallest window of actions where the difference in values is significant. Consider for example the log of a retrieval system is the following:

**Table 1. Local/Global Usefulness Example**

| Search process | Events |
|---|---|
| 1 | enter_search_term→select_term_from_recommender→search →view_record_1→view_record_2→view_record_3→ export_record |
| 2 | enter_search_term→select_term_from_recommender→search →view_record_1→view_record_2→logout |
| 3 | enter_search_term→search→view_record_1→view_record_2 →view_record_3 |
| 4 | enter_search_term→search→view_record_1→view_record_2 →view_record_3→view_record_4→view_record_5 |
| 5 | enter_search_term→select_term_from_recommender→search →view_record_1→view_record_2→bookmark_record→ view_record_3 |
| 6 | enter_search_term→search→view_record_1→export_record |

In the given example (see Table 1) the $LocalUsefulness\,(select\_term\_from\_recommender)$ is then 0.5 as in three out of six search processes the recommender was used.

The $GlobalUsefulness\,(select\_term\_from\_recommender)$ is 0.66 in a window of five succeeding actions as two out of three search processes with term recommender usage are followed by a positive signal in a window of five actions. We can furthermore compare the effect of using and non-using the term recommender on search success. Given this example, the ratio of searches without term recommender usage but incidence of a positive signal is just one out of three (0.33), which differs immensely from the global usefulness of 0.66 found for the term recommender. This result emphasizes the positive effect of the search term recommender on search success, i.e. its usefulness. The values furthermore show that this positive effect correlates not very well with the usage rate (50%) on the local level which may indicate some potential for improvements of the service locally.

A strength of this approach is that it does not need to know the concrete task of the user but appropriate interaction logs. For measuring local usefulness we just need to count the interactions that stand for the service in question. To estimate global usefulness of a service we need to define the set of interactions representing positive signals of search success. This is surely the crucial point of the proposed approach since we need to make some assumptions about positive signals of search success. However, we believe that for each IR system a domain-specific set of positive signals can be defined on the ground of the purpose of the system in question. For the case of a scholarly information portal, for instance, downloading found publications is certainly a strong indication of search success.

In the following we present a case study with a search term recommender provided by a digital library of the Social Sciences where we apply the above introduced measures of local and global usefulness. Our focus in this study is on the occurrence of positive signals in search processes using vs. not using the term recommender.

## 4. EXAMPLE OF USE

### 4.1 Use Case: The Combined Term Suggestion Service

A search term recommenders is a value-added IR service which aims at improving retrieval quality by proposing the user more proper search terms. In [11] we have tested search term recommenders with different vocabularies (user terms, terms from a heterogeneity service, thesaurus terms, co-word analysis) in Sowiport and found that a combination of thesaurus terms and co-word analysis terms works best with respect to user acceptance. The service has been used in about 14% of 3,604 search queries submitted by 1,000 unique users. In this work we build on these results and have implemented an extended recommender service which combines (1) thesaurus terms, (2) additional related thesaurus terms and (3) terms from the Search Term Recommender (STR). The STR works on the basis of co-word analysis from titles and abstracts to thesaurus terms.

The combined search term suggestion service (CTS) [11] is integrated in the Social Science Digital Library Sowiport[1] [10]. The portal contains more than 8 million literature references, full texts and research projects from 18 different databases and reaches more than 20,000 unique users per month. The CTS has been integrated into the search bar on the start page and above the result list for the search form field types "All Fields" and "Keywords". For the other types (Title, Institution, Numbers, Date) we use the autocomplete functionality of the underlying VuFind[2] framework based on the Solr index. On the user interface, in the upper part of the CTS (see Figure 1), users are proposed up to five descriptors from the thesaurus that autocomplete already entered characters. Additionally, for each thesaurus term, all thesaurus terms with a semantic relation like broader, narrower or related are shown in a line underneath in a lighter font color. In the lower part, under the heading "Alternative keywords", the recommender suggests up to five terms from the STR. Figure 1 shows a screenshot of the CTS included in Sowiport.

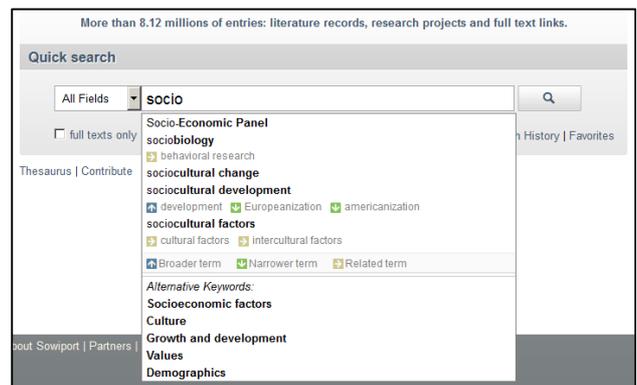

**Figure 1.** A screenshot of the term suggestion service in Sowiport. The service recommends more appropriate terms from controlled vocabulary (such as "Socio-Economic Panel" for the search term "socio") as well as alternative terms returned by a co-word analysis (here, "Socioeconomic factors").

---

[1] http://sowiport.gesis.org

[2] http://vufind.org

### 4.2 Data Set & Methodology

Given our definition of usefulness described above, in this case study we count the frequency to which the CTS has been used (to measure local usefulness) and the degree to which term selections from the CTS co-occurs with positive signals of search success in the following search process. For this, we first need to define the set of positive signals indicating search success. In the case of Sowiport there are a number of interactions on the hit list of a search and in the detailed document view that can be considered as positive indications of a successful search, for example downloading the full text from a record in the result list. See Table 2 for a list of all positive signals and their descriptions.

We then measure the co-occurrence between CTS usages and positive signals on the basis of log data. For this we have used the WHOSE log analysis tool for IIR [9]. The tool allows to load log data from a digital library and to examine user session data with filters, visualizations and a detailed session list with all interactions. We used the tool for the preparation of session data with log data from Sowiport from 15th July 2014 to 15th July 2015 including all user interactions, e.g. a term selection from the CTS by a user is logged by the event CTS_select. See Table 3 for some basic search interactions.

The result is a database that contains all user sessions with its actions and parameters. The tool also prepares session patterns containing the sequence of actions of a session in the form "*action_1>>action_2>>action_3*". To measure local usefulness we used the count of CTS_select occurrences in the data set. To compute the global usefulness of the CTS we measured the co-occurrence between CTS_select and positive signals within a certain event window. For this, we defined a regular expression to identify relationships between CTS_select and positive signals within an event window of *n* actions, for all $n \leq 17$. By this, we obtained both information on the relationship between the occurrence of CTS_select (the use of the term suggestion service) and positive signals of search success as well as information on the point in time (in terms of interactions) in which a positive signal occurs after submitting CTS_select. For the comparison to searches without CTS usage we defined a regular expression to get all event windows with the starting event CTS_search ("CTS usage would be possible…") not preceded by CTS_select ( "…but has not been used"). This makes the comparison between event windows with vs. without CTS usage more precise as no general searches e.g. from internal links or URLs are included.

## 5. RESULTS

### 5.1 Local Usefulness

In the evaluation period 59,568 sessions with 192,024 search queries from search forms with the field type selected to "All Fields" or "Keywords" have been performed, among them 21,448 selected recommendations from the CTS. Figure 2 shows the development of the CTS usage in Sowiport in relation to conducted searches. After the integration of the CTS service in Sowiport in July 2014 in just about 4% of the searches the service has been used. In August/September 2014 the service showed a large increase in usage to 9-10% because of a major speed and cache improvement. Since that time there is a relatively stable usage of around 10% of all searches. Thus, locally (in the query formulation phase) in 10% of searches the CTS service has been considered as useful by users. Without having a reference value or other comparative information it is difficult to judge 10% as a low or a high value. Thus, at first glance, a pure usage rate based notion of usefulness does not appear as a very valuable metric of system quality. However, its benefit opens up if we complement it with global usefulness.

**Table 2. Positive Signals in Sowiport**

| Short | Description |
|---|---|
| goto_fulltext | Follow an external link which leads to a full text in PDF or HTML format |
| goto_google_scholar | Search the record in Google Scholar |
| goto_google_books | Search the record in Google Books |
| goto_local_availability | Check for availability in a local library |
| view_description | View the record's abstract |
| view_citation | View the record's citation data |
| view_references | View the record's references |
| export_cite | View record in different citation styles |
| export_bib | Export the record to different citation formats |
| export_mail | Send the record via email |
| save_to_multiple_favorites | Check several records in the result list and save them to favorites |
| to_favorites | Save a single record to favorites |
| export_search_mail, | Send the search via email |
| save_search | Save the search to favorites |
| save_search_history | Save a search from the history to favorites |

**Table 3. Some basic search interactions in Sowiport**

| Short | Description |
|---|---|
| CTS_select | A user selects a term from the CTS |
| CTS_search | A search from a search form with the field type selected to "All Fields" (default setting) or "Keywords" |
| search | A general search from a search form, an internal link or by URL from a search engine etc. |
| view_record | A click on a record in the result list to see the detailed view of a record |

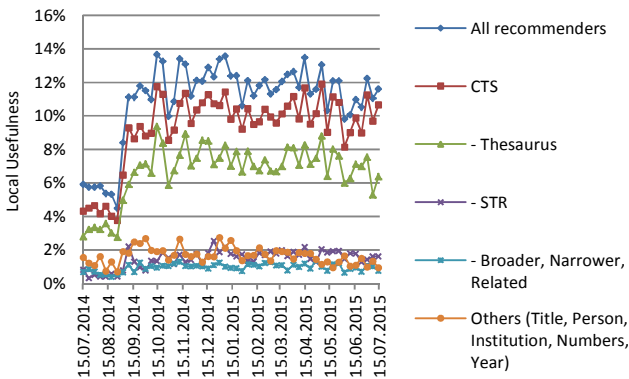

**Fig 2.** Recommender usage in Sowiport in percent of all conducted searches

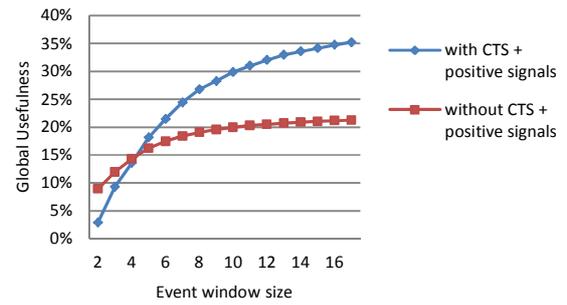

**Fig 5.** The global usefulness of searches with vs. without use of a combined term suggestion service (CTS)

### 5.2 Global Usefulness

To analyze the global usefulness of the CTS we follow the evaluation methodology from Section 3. Figure 3 gives an overview of extracted patterns for CTS_select followed by at least one positive signal within an event window of seven succeeding actions (4,569 sessions). Positive signals are color-coded in green. The main pattern which leads to a positive signal is: "CTS_select>>CTS_search>>(view_record)+>>{positive_signal}". Figure 4 then shows the analog diagram for searches without the use of CTS_select in advance and positive signals within an event window of seven actions (21,712 sessions). Here, the main starting point is CTS_search (one event less than for CTS_select). Figure 3 and 4 show that the main path patterns differ not very much between the two search variants. However in Figure 5 it can be seen that searches with CTS usage lead much more frequently to positive signals than searches without (statistically significant for *window size≥5* with Chi-Squared-Test, *p<0.001*). About 14% of the searches lead to positive signals after four interactions, independently of having CTS used before or not. Beginning with step 5, however, the amount of CTS usage followed by at least one positive signal differs significantly from searches without the usage of CTS. The success rate of searches where the CTS has been used increases to a value of 30% (after 10 interactions, and increases further to a value of around 35%) whereas searches where the CTS has not been used achieve a value of 20% (after 10 interactions and do increase only slightly). Within an event window of seven interactions searches *with* CTS usage achieved a global usefulness of 0.24 whereas searches *without* CTS usage achieved a global usefulness of 0.18. This is clearly a significant difference in favor of the CTS service. The CTS service improves the search success at a rate of about 20-35%. Thus, the CTS service seems to be – globally – a useful service since it shows a high potential in increasing retrieval quality in terms of search success. The interesting finding now is that from the perspective of the user the "true" usefulness of the CTS service is not evident at the local level, in the moment when the user has to make a choice of using or leaving the search term recommender. Its benefit becomes apparent in a later phase of the session. This discrepancy may provide a clue to system developers to improve transparency of the service at the local level (e.g. by providing a preview of search results).

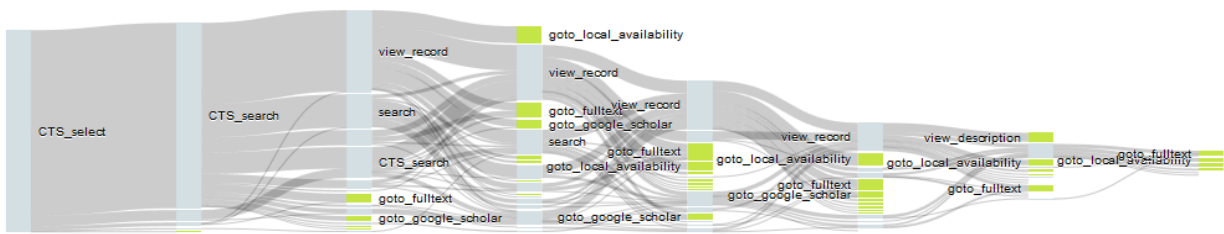

**Fig 3.** Path patterns for searches with the use of the CTS + positive signals (green colored) within an event window of seven actions (Node labels for p>0.02 and positive signals with p>0.005)

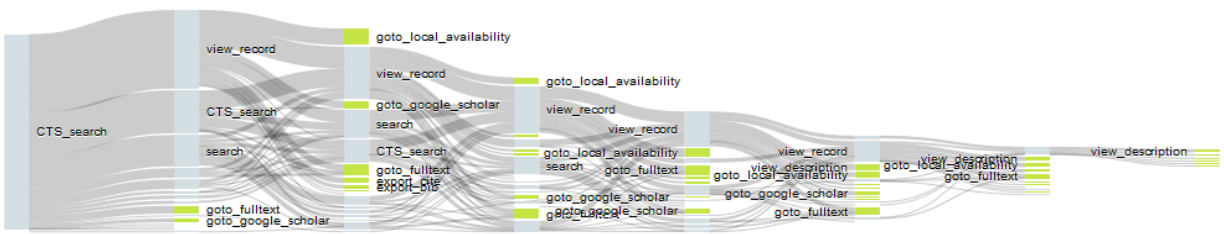

**Fig 4.** Path patterns for searches without the use of CTS + positive signals (green colored) within an event window of seven actions (Node labels for p>0.02 and positive signals with p>0.005)

**Table 4. Mean P@20 and MAP@20 for different session units. An asterisk (*) indicates that the means are significantly different with a two-sample Z-Test and p<0.0001. The numbers in parentheses show the average count of searches per session unit.**

| Session unit | Mean P@20 without CTS | Mean P@20 with CTS | MAP@20 without CTS | MAP@20 with CTS |
|---|---|---|---|---|
| All whole sessions | 0.0685 (9.5) | 0.0677 (13.1) | 0.1177 | 0.1112 |
| Sessions with searches > 1 click-through signal | 0.1853 (3.5) * | 0.1886 (4.7) * | 0.3183 | 0.3099 |
| Succeeding search process | 0.0624 (1.0) | 0.0651 (1.0) | 0.1904 * | 0.1692 * |
| Window (7) - split to single searches - Click-through signals | 0.0515 (1.5) * | 0.0476 (1.9) * | 0.1550 | 0.1403 |
| Window (7) - split to single searches - All positive signals | 0.0562 (1.5) * | 0.0513 (1.9) * | - | - |
| Window (7) - split to single searches - with searches >1 signal - Click-through signals | 0.1478 (0.5) * | 0.1371 (0.7) * | 0.4452 * | 0.4040 * |
| Window (7) - split to single searches - with searches >1 signal - All positive signals | 0.1599 (0.5) * | 0.1465 (0.7) * | - | - |
| Window (7) - no split to single searches - Click-through signals | 0.0709 (1.0) * | 0.0869 (1.0) * | 0.2062 * | 0.2440 * |
| Window (7) - no split to single searches - All positive signals | 0.0779 (1.0) * | 0.0941 (1.0) * | - | - |

# 6. COMPARISON TO STANDARD IR MEASURES

In the following we compare the results for global usefulness from the previous section to traditional IR measures such as precision (P@k) and mean average precision (MAP). The underlying research question addresses the difference between sessions with and without CTS usage (analog to the results for global usefulness) to measure the usefulness of the CTS service.

Thereby, we evaluate both precision measures with variable parts: (1) the *session unit* to which the measure is applied to. For different *session units* we use the *whole session* itself, the *succeeding search process* after the use of an IR service and a growing *window of actions* after the usage of the IR service.

(2) the *type of signals* the measures use. For global usefulness we have used the set of positive signals listed in Table 2. For precision measures relevance judgements for document relevancy from real users are needed as ground truth. However, in the case of log-based evaluation these are most times missing. There are several ways to recover them: (1) referring to an external source of user judgments, (2) asking a set of users subsequently to rate document relevancy for topics or (3) click-through data as implicit user feedback for the relevancy of documents can be used. In the following we use different signals for implicit user feedback.

Precision measures describe the number of relevant documents in a search result. P@k takes only the first *k* positions into account. Mean precision (MP) takes the mean of precision values for multiple search results. Average precision (AP) additionally takes the position of a relevant document into account. Mean average precision (MAP) then takes the mean of average precision values for multiple search results.

The result list in Sowiport shows twenty documents. Accordingly, we use P@20 as a first precision measure and MAP@20 as a second measure. For P@20 we use five actions on the result page for *click-through signals*: click on the title for a detailed view of the record (view_record, see Table 3), click on the full text link, click on the Google Scholar link, click on the Google Books link, and adding the document to favorites. Thus, the set of a*ll positive signals* contains the positive signals from Table 2 plus the view_record signal. For MAP we use only the click to the detailed view as all other actions have no rank information in our logs. All click-through signals appear on the initial result page directly after the search commit and are directly connected to a single document. It should be noted that these signals surely provide no reliable indications of relevance, but some clues of search success in terms of getting the user to start a more detailed inspection of the information retrieved. Thus, we just apply precision measures to the incidence of positive signals among the top 20 documents.

Table 4 shows the results for different session units. The first row shows the mean P@20 and MAP@20 over *all whole sessions* with and without CTS usage. The sessions were split into single search processes. Then, in each search process we look for click-through signals. As one can see in the table the precision and MAP values are relatively low. A deeper look into the data showed a very high number of query reformulations and actions on the result list such as filtering with facets before looking at a document in detail. Theoretically, for session-based mean precision measures all query reformulations need to be merged to a topic and all clicked documents over the session need to be assigned to the topics. This forms an own complex problem which has been already addressed in research [e.g. 19]. As a naïve approach we computed the sessions' precision only with search processes which include at least one click-through signal (see Table 4, second row). Here, the precision values increase strongly. For P@20 the difference between searches with CTS usage and searches without CTS usage are statistically significant but the absolute difference is still very low. The numbers in parentheses show the average count of searches per session unit. By comparison to the first row we find that only every third search process includes at least one click-through signal. This is a further indication for a high number of query reformulations.

As already mentioned, precision and average precision are related to a single search result. We accordingly tested the precision of the *search process* which directly follows the events CTS_search and CTS_select. The table's third row shows that the results are again very low for P@20 and low for MAP@20. Here, the same problem as in sessions as a unit occurs: the first query after submitting from a search form is often only the starting point for a set of several query reformulations and search adaptions.

Similarly to our procedure for global usefulness we then tested with the *event window approach*, which takes the first seven succeeding actions after submitting a search (either with or without CTS usage). This window has then been *split to single searches*. The window involves on average 1.9 searches for windows with CTS usage and 1.5 searches for windows without CTS usage. We can see in the fourth and fifth row that the mean precision values are still very low. By counting again only searches with more than one following signal the precision increases strongly. Finally, taking all positive signals into account the precision increases additionally by about 1%. So far, with the window-based approach and splitting to single searches we found statistical differences (mainly because of the high sample size), but the precision values are very low. Also, the number of searches per window are different (1.9 searches for CTS_select, 1.5 for Search). This makes it hard to compare the two kinds of sessions.

Therefore, in the next approach we did *not split the window to single searches*, but look the whole window as a single search. This means, we count all signals inside the window and do not assign the signals to single searches which will give us a different view. The last two rows in the table show that there is a significant difference for both P@20 and MAP@20 in favor of sessions with CTS usage.

To understand how the MAP measure evolves over increasing window sizes we show the precision graphs in Figure 6. It shows MAP@20 with a window of growing size, once for the window split to searches, and once not split to searches. The curves for the "with CTS split" remain at 0.14, for "without CTS split" slightly higher at 0.15. The curves for the no-split window diverge. Differences between MAP@20 for "with CTS no-split" and "without no-split" are significant for a window size≥6 with a two-sample Z-Test and p<0.0001. Here it can be seen that with the window-based approach and not splitting the window to single searches the differences between sessions with and without CTS usage are getting larger by increasing the window size. This means for MAP that significantly more top-ranked documents are viewed after CTS usage than without it and the effect becomes larger in the following steps of the session. If we normalize the effect of more searches per window for CTS sessions not split to searches, the difference between sessions with and without CTS usage becomes really measurable also with the MAP measure.

## 7. DISCUSSION

The case study showed a local usefulness of 10% for the CTS. When we contrast this value with the global version of usefulness it turned out that the local usage of the service under study does not correspond with its positive effect on search success (in terms of positive signals). Thus, the "true" usefulness of the CTS service becomes apparent in a later phase of the session, but is not shown to the user at the local level. Usefulness therefore is a concept that obviously needs both a local and a global version contrasting the effect of a local ISS to the global benefit of the system in helping the user accomplishing his/her information seeking goal.

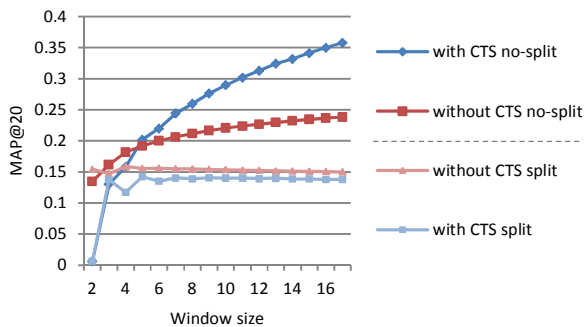

**Fig 6.** MAP@20 for a window of succeeding actions with growing size. One time the window has been split to searches (split), the other time not (no-split).

## 7.1 Signals and their variations

A critical point of our approach is surely the definition of a set of positive signals. In our case study positive signals are either bounded to search results (email, save, RSS) or documents (download full text, view abstract, save, export etc.). Positive signals on the document level can be an approximation of document relevance in the case where explicit relevance assessments are not available. This way, relevance can be computed with a log file-based approach. However, in other use cases probably positive signals exist which are bounded to different artefacts of the search process (such as query formulation, results scanning, working with favorite documents and searches, browsing in the document space etc.). Choosing different categories of positive signals can adopt the use case, but also makes the usefulness measure applicable to the whole search process. Another factor is the weighting of signals. In this case study we weighted every signal equally. However, one could argue that exporting or saving a record weights more than just viewing the abstract or its references. Here, definitely more research in the field of (positive) signals in the search and browsing process needs to be carried out.

In our study we utilized actions from log files as indications for positive signals. Research showed (compare Section 2.3) that more complex actions such as a record view with a certain dwell time represent well document usefulness or user satisfaction. Additionally, there are also signals from other eye-, mouse- or keyboard tracking that can identify user satisfaction. There is no difficulty to use these signals in our approach, however signals may vary from system to system, may not be available in every log and computability could increase.

Depending on the type of the IR Service there can also be a set of signals which indicates a negative impact of the service, for example, by a quick exit of the session or by dropping a document from the favorite list. An obvious solution to this problem is to compute global usefulness for these "negative" sessions separately and then subtract it from the global usefulness of the "positive" sessions.

## 7.2 Path patterns

A deeper look into path patterns might reveal insight which typical action sequences lead to positive signals. In our case searches seem to follow a relatively straightforward pattern, such as CTS_select>>CTS_search>>view_record>>{positive_signal}. This observation raises further research questions: "Are there any regularities in search path patterns?", "Causes the use of particular IR services differences in patterns?", "Are there any differences between different kind of users?", and so on. More insights as regards search patterns can reveal a ground for optimization of IR services.

## 7.3 Differences to standard IR measures

Global usefulness measures the observation of at least one positive signal (like bookmarking a record) within a fixed window of actions after the use of an IR service (like a search term recommender). In contrast, P@k and MAP measure the quality of a search result by the presence and ranking of relevant documents. Thus, there are a number of differences between global usefulness and classical measures such as precision or MAP: (1) Global usefulness measures *success* of an IR service in terms of occurrence of positive interaction signals, P@k/MAP measures the *quality* of a search result in terms of relevance; (2) Global usefulness evaluates interactions in a fixed window of actions whereas P@k/MAP evaluates a specific search result. (3) Global usefulness uses positive signals obtained from log data whereby P@k/MAP uses relevance assessments usually obtained from human experts. In Section 6.1 we have analyzed these differences by substituting relevance assessments by click-through signals and applying different session units (wrt 2.) and different signals (wrt 3.) to the precision measures.

(Regarding 2.) Precision measures are per definition bound to a single search result. Because in our data set we found a high number of query reformulations and adaptions this leads to low mean values because not in every search process users clicked on documents. The succeeding search directly after the usage of the IR service also seems to be inappropriate as it is only the starting point for a number of reformulations. We then choose a window approach which allows to go further into the sessions and captures about 1.9 search processes for windows with CTS usage in a window of 7 actions. But here again the number of searches keeps the precision values on the same low level and significant differences between sessions with or without CTS usage are difficult to prove. Not splitting the window to single searches then shows a strong increase in precision and MAP. Additionally, a significant difference between sessions with/without CTS usage appears to be similar to the results for global usefulness.

(Regarding 3.) Global usefulness uses a set of positive signals (without the view_record signal) whereby MAP uses only the view_record signal. However, both lead to significant differences for sessions with and without CTS usage. This seems to measure the same but it does not: the CTS usage led to more relevant documents (MAP) and these subsequently to more positive signals (global usefulness). However, global usefulness can measure also signals apart from document relevancy, e.g. when a user exports the whole search result.

## 8. CONCLUSION

In this paper we propose a specific log data based methodology that measures how useful an IR service is for the user. Corresponding to the model of Cole et al. [4] usefulness is measured on several levels. In our approach we distinguish between local and global usefulness of the IR service. Local usefulness asks for "the systems support toward the goal of each interaction", i.e. we measure how often the IR service is used at the local level and which patterns of usage occur. Our case study turned out that the service under study (a combined term suggestion service (CTS)) was used in about 10% of cases. On the level of global usefulness Cole et al. ask for the contribution of each information interaction to accomplishing the sub goal and the overall task or goal. Since the user's task is often difficult to capture ouu approach is to approximate usefulness by looking at positive signals in the search process and investigating how well the search service's local usage co-occurs with positive signals. Our case study showed that CTS usage has a significantly stronger relationship with positive signals than searches without it.

In general, our attempt can contribute to a new approach of measuring usefulness of IR supporting services with regard not only to its usage, usability or its influence on search results but encompassing the whole search process. Unlike user-oriented studies our approach is based on log files and scales well on thousands of users and sessions. Nevertheless, the users will not be lost of sight, as their individual behavior is stored in the log files.

The work done in this paper is insofar only theoretical, as we take positive signals in search sessions as indications of user's preference. A critical next step is to compare our findings and measures from this work with insights from user studies where we intend to compare user feedback on usefulness with log-based findings.

### ACKNOWLEDGMENTS
This work was partly funded by the DFG, grant no. MA 3964/5-1; the AMUR project at GESIS. The authors thank the focus group IIR at GESIS for fruitful discussions and suggestions.